\documentclass[12pt]{article}
\usepackage{epsf}
\usepackage{a4wide,graphicx}
\usepackage{cite}

\newcommand{\mev}{\textrm{ MeV}}
\newcommand{\be}{\begin{equation}}
\newcommand{\ee}{\end{equation}}
\newcommand{\ba}{\begin{eqnarray}}
\newcommand{\ea}{\end{eqnarray}}

\begin{document}

\title{Radiative decay into $\gamma P$ of the low lying axial-vector mesons}

\author{
H. Nagahiro$^1$, L.~Roca$^2$ and E.~Oset$^3$\\
{\small{\it $^1$Research Center for Nuclear Physics (RCNP), Ibaraki, Osaka
  567-0047, Japan}}\\
{\small{\it $^2$Departamento de F\'{\i}sica. Universidad de Murcia. E-30071
Murcia.  Spain}}\\
{\small{\it $^3$Departamento de F\'{\i}sica Te\'orica and IFIC,
Centro Mixto Universidad de Valencia-CSIC,}}\\
{\small{\it Institutos de
Investigaci\'on de Paterna, Aptdo. 22085, 46071 Valencia, Spain}}
}

\date{\today}

\maketitle

 \begin{abstract} 
 
We evaluate the radiative decay into a pseudoscalar meson and a photon of
the whole set of the axial-vector mesons dynamically generated from the
vector-pseudoscalar meson ($VP$) interaction. We take into account tree level
and loop diagrams coming from the axial-vector decay into a vector and a
pseudoscalar meson. We find a large span for the values of the radiative widths
of the different axial-vector mesons. In particular, we  evaluate the
radiative decay into $K \gamma$ of the two $K_1(1270)$ states, recently
claimed theoretically, and discuss the experimental values quoted  so far on
the assumption of only one state.

\end{abstract}

\section{Introduction}
The radiative decay of resonances has always been one of the basic observables 
providing insight into the nature of the states. Within quark models it has
been thoroughly investigated, concerning mostly the radiative decay of baryon
resonances \cite{capstick,cauteren,umino,Myhrer:2006cu}. Regarding axial-vector
mesons, the $a_1^+$ radiative decay has 
been studied within different contexts, for instance vector meson
dominance is used in \cite{shuryak,haglin}, relating the radiative
decay with the $\rho \pi$ decay of the $a_1^+$.  Chiral Lagrangians
with vector meson dominance (VMD) are also used in
\cite{Ecker:1988te} to obtain the radiative width of $a_1^+ \to
\pi^+\gamma$. The rates of $a_1^+ \to \pi^+\gamma$ and $b_1^+ \to \pi^+\gamma$ 
are also
evaluated in  \cite{rosner} using quark
 models, or effective Lagrangians \cite{Roca:2003uk},
 for the $a_1 \to \pi \rho$ and 
$b_1 \to \pi \omega$ and VMD to relate these amplitudes with the radiative
decay. 

A new approach is required
for the resonances which qualify as dynamically generated from the
meson-meson
or meson-baryon interaction.  This is so because, being the meson or baryon
components the basic building blocks, the decay into meson photon or baryon
photon is obtained by coupling the photons to the meson or baryon components of
the resonance.  In this direction the radiative decay of the $\Lambda(1520)$
has been recently studied \cite{Doring:2006ub}, as well as that of the $\Delta(1700)$
\cite{Doring:2007rz}.  Concerning the radiative decay of 
axial-vector mesons, work in this direction
has also been done in \cite{Roca:2006am}  evaluating the radiative widths of the 
$a^+_1(1260)$ and  $b^+_1(1235)$.  The $a_1$ and $b_1$ axial-vector mesons are
part of the two SU(3) octets and one singlet states which are dynamically
generated from the interaction of vector mesons with pseudoscalar mesons. By
using chiral Lagrangians and techniques of chiral unitary theory one constructs
the s-wave scattering amplitudes for vector-pseudoscalar in coupled channels
and looks for resonances either using the speed plot \cite{kolo}, or searching
for poles in the second Riemann sheet \cite{Roca:2005nm}. Several states appear which
can be associated to the $h_1(1170)$, $h_1(1380)$, $f_1(1285)$, $a_1(1260)$,
$b_1(1235)$, $K_1(1270)$ resonances. In \cite{Roca:2005nm} two poles for the $K_1(1270)$
resonance were found, in analogy with the two poles found for the
$\Lambda(1405)$ \cite{jido,carmina,hyodo}, for which experimental evidence has 
been found in \cite{magas} from the analysis of the 
$K^- p \to \pi^0 \pi^0 \Sigma^0$ reaction of \cite{prakhov}. In a similar way,
experimental support for the two $K_1(1270)$ states has been recently shown in
\cite{Geng:2006yb}.

   From this perspective we consider all the low lying axial-vector
meson states mentioned above and evaluate their radiative
decay width for the different charge states. Among others, we look
now at the radiative decay of the neutral $a_1$ and $b_1$ states to
complement the evaluations done before in \cite{Roca:2006am} for
the charged states. 

   The experimental situation is not very rich, something that should be
reversed now that we are finding new motivations for more data. Apart
from the information on the charged $a_1$ and $b_1$ decay widths, there is only
information on the radiative decay width of the neutral $K_1(1270)$ state
obtained with Primakoff scattering of $K_L$ with nuclei at high energies 
\cite{Alavi-Harati:2001ic}.  
We argue that the existence of the two $K_1(1270)$ states
 blurs the
conclusions obtained for this width in \cite{Alavi-Harati:2001ic},
 since some of the assumptions 
made to extract this number would require a revision 
after the findings of
\cite{Roca:2005nm} and \cite{Geng:2006yb}. We make predictions in the paper for the decay
widths of all these resonances and make suggestions of experiments to further 
support the existence of the two $K_1(1270)$ states.

\section{Summary of the formalism}

In this section we briefly summarize the formalism described in
ref.~\cite{Roca:2006am} for the evaluation of the $b_1^+ \to
\pi^+\gamma$ and $a_1^+ \to \pi^+\gamma$ decays and generalize the
model to the other axial-vector mesons mentioned above.

\begin{figure}[hbt]
\epsfxsize=12cm
\centerline{
\epsfbox{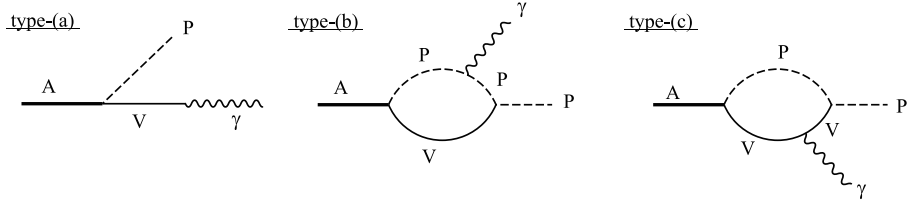}}
\caption{Feynman diagrams needed in the evaluation of
the radiative axial-vector
 meson (A) decay.}
 \label{fig:feynman}
\end{figure}

In ref.~\cite{Roca:2005nm} it was shown that, with the
implementation of unitary techniques in the evaluation of the
s-wave  scattering amplitude for the interaction of the octet of
vector (V) mesons and the octet of pseudoscalar (P) mesons, many of
the low-lying axial-vector resonances show up as poles in
unphysical Riemann sheets of the unitarized $VP$ amplitudes.
Therefore, these resonances qualify as dynamically generated.
In view of the dominant contribution of the $VP$ channels in the
building up and decay of the axial-vector resonances, the
philosophy to calculate the radiative decay is to consider the
transition of these resonances to the allowed VP channels, either
at tree level and one loop, and attach the photon to the allowed
meson lines and vertices, see fig.~\ref{fig:feynman}.

\begin{figure}[hbt]
\epsfxsize=0.9\textwidth
\centerline{
\epsfbox{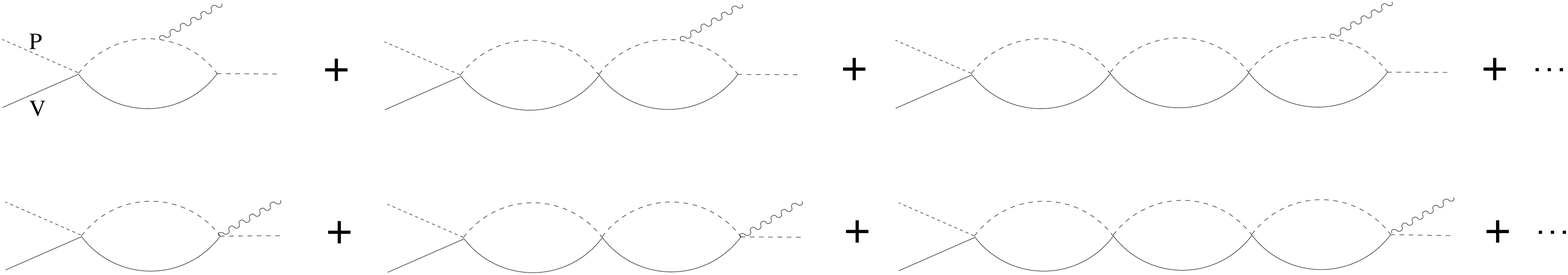}}
\caption{Series implicit in the type-b loop of fig.~\ref{fig:feynman} in
the dynamically generated picture of the axial-vector resonances}
 \label{fig:seriesnew}
\end{figure}

\begin{figure}[hbt]
\epsfxsize=0.65\textwidth
\centerline{
\epsfbox{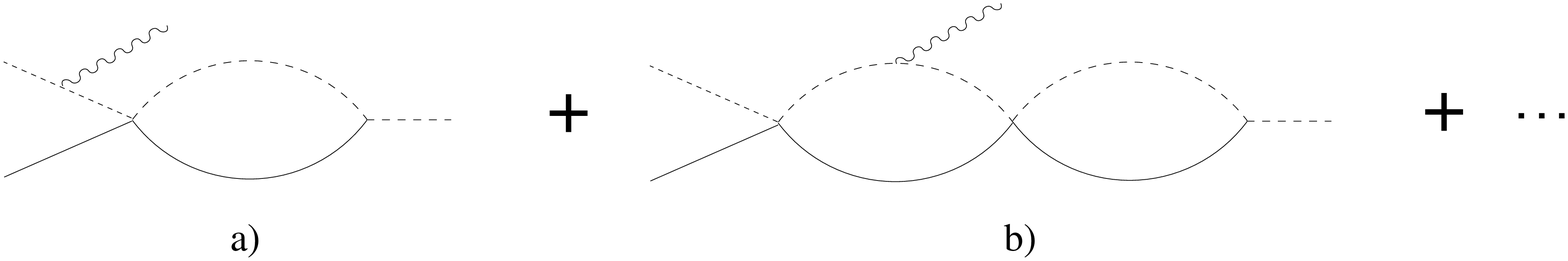}}
\caption{Extra allowed diagrams required by gauge invariance.}
 \label{fig:seriesbnew}
\end{figure}

\begin{figure}[hbt]
\epsfxsize=0.65\textwidth
\centerline{
\epsfbox{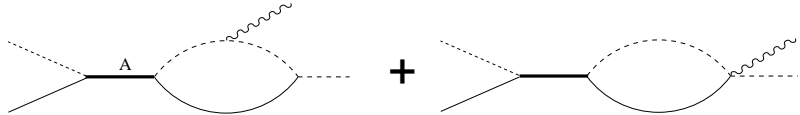}}
\caption{Equivalent representation of fig.~\ref{fig:seriesnew}
close to the axial-vector meson pole position.}
 \label{fig:4new}
\end{figure}

\begin{figure}[hbt]
\epsfxsize=0.85\textwidth
\centerline{
\epsfbox{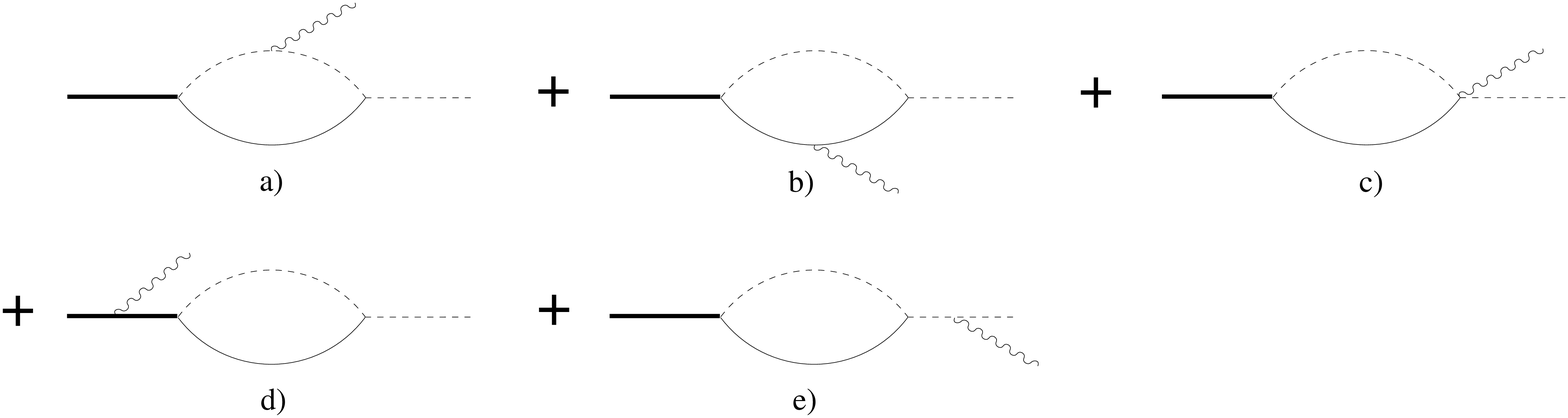}}
\caption{Set of diagrams needed {\it a priori} in the
evaluation of the axial-vector meson radiative decays.}
 \label{fig:5new}
\end{figure}

In ref.~\cite{Roca:2006am} it was shown that, by invoking gauge
invariance, only these diagrams need to be evaluated.
In the next paragraph we elaborate further on this point.
Let us look at the loop diagrams of fig.~\ref{fig:feynman}.
Keeping in mind the dynamical origin of the resonance from the
Bethe-Salpeter resummation of loops containing the kernel of the
$VP\to VP$ interaction, the series implicit in those loops is given
in fig.~\ref{fig:seriesnew}, where we have also added the second
raw of diagrams to be discussed later on.  
The photon coupled to the vector in the loop should be understood
in the discussion, but is omitted to save diagrams.
The requirement of gauge invariance would demand that
the photon  couples to all lines in the loops and vertices.
This has been done in several works 
\cite{Nacher:1999ni,Borasoy:2005zg,Borasoy:2007ku}
dealing with photonuclear processes which involve dynamically
generated resonances. An explicit proof of gauge invariance of
this kind of diagrams can be seen in \cite{Borasoy:2005zg}. Thus,
in addition  to the diagrams of fig.~\ref{fig:seriesnew} we would
have diagrams like those in fig.~\ref{fig:seriesbnew}.
Note that the $VP \to VP$ vertex is of the type
$\epsilon_V\cdot\epsilon_V'$ \cite{Roca:2005nm}, 
with $\epsilon_V$ and 
$\epsilon_V'$ the polarization vectors of the vector mesons, and thus has not
photon contact term associated of the type $VVPP\gamma$. Diagrams
$a)$ and $b)$
of fig.~\ref{fig:seriesbnew} are proportional to the
last loop function  with intermediate $P$ and $V$, which has the
structure $J(Q^2)Q^\mu$, with $Q$ the momentum of the produced
pseudoscalar. However, as shown in the appendix of 
\cite{Roca:2006am}, this loop function satisfies $J(Q^2=m_P^2)=0$,
where $m_P$ is the mass of the produced pseudoscalar in the decay
(see also ref.~\cite{Gamermann:2007bm} for an alternative derivation
with standard vector mesons, not dynamically generated). This is
due to the requirement that the longitudinal part of the
axial-vector propagator must not develop a pole of the pseudoscalar
\cite{Kaloshin:1996kz}. Radiation from the final pseudoscalar (see
fig.~\ref{fig:5new}) also leads to a null contribution, as discussed
later. Thus, we are left with the diagrams of
fig.~\ref{fig:seriesnew} where the photon couples to the last loop,
from where the pseudoscalar is emitted. The sum of loops before the
last one generates the $VP$ T-matrix that contains the pole for the
axial-vector \cite{Roca:2005nm}. The sum of diagrams is thus
equivalent to the diagrams of fig.~\ref{fig:4new}.
The sum of diagrams of fig.~\ref{fig:seriesnew} lead to a
$VP\to\gamma P$ amplitude, in a simplified way omitting polarization
vectors for simplicity, 
\begin{equation}
-it=-it_{VP\to VP} L
\end{equation} 
where $L$ stands for the last loop function. Since $t_{VP\to VP}$
contains the axial-vector pole, close to the pole position,
$s_p\simeq M_A^2-iM_A\Gamma$, we have 
\begin{equation}
-it_{VP\to VP}=-i\frac{g_{AVP}^2}{s-s_P}.
\end{equation} 
Alternatively, from fig.~\ref{fig:4new} we would have
\begin{equation}
-it_{VP\to P\gamma}=-ig_{AVP}\frac{i}{s-s_P}(-i)g_{AP\gamma},
\end{equation} 
from where
\begin{equation}
g_{AP\gamma}=g_{AVP} L,
\end{equation} 
which is what we would directly obtain from the evaluations of
diagrams of fig.~\ref{fig:feynman} and what is done in 
\cite{Roca:2006am}, and which is the formalism followed in the
present paper.
The former slightly simplified derivation can be followed with more detail
in
\cite{Doring:2006ub} in the study of the radiative decay of the
$\Lambda(1520)$ resonance.
Once the equivalence of these formalisms is established, we can go
back and reconsider the terms that one would have, which are shown
in fig.~\ref{fig:5new},
the last two diagrams contributing in principle for charged
axial-vector states. An explicit proof
of the gauge invariance  and
finiteness of the set of diagrams of fig.~\ref{fig:5new}, for the analogous case of
$P\to V\gamma$ in the charm sector,  is
provided in \cite{Faessler:2007gv}. Furthermore, in the appendix of
\cite{Roca:2006am} and \cite{Gamermann:2007bm} it is shown that
diagram $e)$ of fig.~\ref{fig:5new} vanishes due to the Lorenz
condition of the axial-vector meson ($\epsilon_A\cdot P_A=0$) (also
noted in \cite{Faessler:2007gv}) and diagram $d)$
of fig.~\ref{fig:5new} vanishes due to the condition
$J(Q^2=m_P^2)=0$, required to avoid a pole of the pseudoscalar in
the longitudinal part of the axial-vector propagator
\cite{Kaloshin:1996kz}, as stated above when discussing 
fig.~\ref{fig:seriesbnew}.

After the above discussion we briefly recall the procedure
followed in \cite{Roca:2006am} to evaluate the radiative width,
making explicit use of gauge invariance which simplifies
considerably the calculation.
Since the only external momenta available are 
 $P$ (the axial-vector meson momentum)  and $k$
  (the photon momentum), 
 the general expression of the amplitude can be written as 
  \be
 T={\epsilon_A}_\mu \epsilon_\nu T^{\mu\nu}
 \label{eq:Tgeneral}
 \ee 
with
\be
T^{\mu\nu}=a\, g^{\mu\nu} + b\, P^\mu P^\nu + c\, P^\mu k^\nu
 +d\, k^\mu P^\nu + e\, k^\mu k^\nu
 \label{eq:Tmunuterms}
\ee
In Eq.~(\ref{eq:Tgeneral}),
$\epsilon_A$ and $\epsilon$ are the axial-vector 
meson and
photon polarization vectors respectively.
Note that, due to the Lorentz condition, ${\epsilon_A}_\mu P^\mu=0$,
${\epsilon}_\nu k^\nu=0$, all the terms in Eq.~(\ref{eq:Tgeneral})
vanish except for the $a$ and $d$ terms. On the other hand, gauge
invariance implies that $T^{\mu\nu}k_\nu=0$, from where one gets
\be
a=-d\,P\cdot k.
\label{eq:ad}
\ee
This is obviously valid in any reference frame, however, in the
axial-vector meson rest frame and taking the Coulomb gauge for the
photon, only the $a$ term survives in  Eq.~(\ref{eq:Tgeneral})
since  $\vec P=0$ and $\epsilon^0=0$. This means that, in the end,
we will only need the $a$ coefficient for the evaluation of the
process. However, the $a$ coefficient can be evaluated from the $d$
term thanks to Eq.~(\ref{eq:ad}). The advantage to evaluate only
the $d$ coefficient is that the contact term of
fig.~\ref{fig:5new}c) does not contribute to the $d$ coefficient,
only the loop diagrams of fig.~\ref{fig:feynman} contribute, and
from dimensional reasons (performing explicitly the Feynman
integrals) one can see that the $d$ coefficients are finite for the
diagrams of type-b in fig.~\ref{fig:feynman}. For the type-c
diagrams, as discussed in \cite{Roca:2006am} (after  Eq.~(25)),
there was formally a logarithmic divergence coming from the
$1/M_V^2$ term of the vector meson propagator, which required some
tadpole from higher order terms for cancellation (see also
ref.~\cite{Napsuciale:2007wp} in the analogous problem of
$e^+ e^-\to\phi f_0(980)$). 
In order to evaluate it we must use some regularization procedure.
The most appropriate way to regularize the $1/M_V^2$ terms  is to
connect the divergences with those appearing in the basic problem
of $VP\to VP$ scattering \cite{Roca:2005nm}.
 These divergencies already appear in the loop
containing one pseudoscalar and one vector meson, which was regularized in
\cite{Roca:2005nm} making use of the N/D method of
\cite{Oller:1998zr} and dispersion relations. These allowed one to
factorize on shell terms appearing in the numerator of the loop
functions like the $q^2/M_V^2$ terms, with $q$ the vector meson
momentum of the $VP$ loop \cite{Roca:2005nm}. In the present case
one can estimate the contribution of the $q^2/M_V^2$ or the type-c
loop by realizing that if one looks for sources of imaginary part
by cutting diagram c) of fig.~\ref{fig:feynman} by vertical lines,
the cut to the left of the photon line can place two mesons on
shell, for instance $\pi$ and $\rho$ for the $a_1$ resonance. The
cut to the right of the photon, which would correspond to the
energetically forbidden $\pi \to \pi \rho$ decay, does not provide
imaginary part. Thus, the only source of the imaginary part comes
from the cut at the left of the photon line, which is
 the same as
that for the basic pseudoscalar-vector meson loop of the scattering
problem. This allows us to replace momentum 
factors appearing in the
numerator of the loop function, {\it i.e.} factors $q^2/M_V^2$,
by its on shell value.
The substitution is done in \cite{Roca:2005nm} after the $q^0$
integration is performed and one replaces $q^2/ M_V^2$ by
$\vec{q}\,^2_{\textrm{on}}/M_V^2$.  Hence, the effects here are also of the order
of $\vec{q}\,^2_{\textrm{on}}/M_V^2$ (there was an extra factor 1/3
for symmetry reasons in \cite{Roca:2005nm}, which we ignore here to
give a conservative estimate of the effects). Hence, we have
checked that, for the most relevant cases, the estimate gives an
upper limit of the order of 20\% of the rest of the c) diagram.
This value is of the same size as  the finite results found in the
diagrams of type-b, where the effect of the $1/M_V^2$ terms was of
the order of $10\%$ or less.   In the final results we will add in
quadrature to the theoretical uncertainty a very conservative  10\%
of the total radiative decay width   from this neglect of the
$1/M_V^2$ terms in the type-c loops. \\

The Lagrangians needed in the evaluation of the diagrams 
in fig.~\ref{fig:feynman} are given in ref.~\cite{Roca:2006am}.
From these Lagrangians the tree level amplitude, type-a in
 fig.~\ref{fig:feynman}, takes the form
 
\begin{equation}
t_a = - g'_{AVP} e \lambda_V F_V \frac{1}{M_V}
 \epsilon_A\cdot\epsilon
\label{eq:tree}
\end{equation}
with $\lambda_V=1$, $1/3$, $-\sqrt{2}/3$ for $\rho$, $\omega$ and
$\phi$ respectively, $F_V=156\pm 5$MeV \cite{Palomar:2003rb},
 $M_V$ is the vector
meson mass and $e$ is taken 
positive.
In Eq.~(\ref{eq:tree}), $g'_{AVP}$ is
 the $AVP$ coupling in the charge base. These
 coefficients  are related to the
 $g_{AVP}$
 in isospin base, obtained in
ref.~\cite{Roca:2005nm},  through the transformation
\begin{equation}
g'_{AVP} = {\cal C} \times g_{AVP},
\label{eq:gAVP}
\end{equation}
where ${\cal C}$ are coefficients dependent on the different 
$AVP$ channels 
summarized in tables~\ref{tab:h1}--\ref{tab:K0}.
In the present work we use the values of $g_{AVP}$ obtained in
refs.~\cite{Roca:2005nm,Roca:2006am} by evaluating the residua at
the pole position of the different $VP\to VP$ scattering
amplitudes.\\


\begin{table}[h]
\begin{center}
\begin{tabular}{rc|r|r|r} 
\multicolumn{2}{c|}{$h_1(1170/1380)\rightarrow\pi^0\gamma$} & ${\cal C}$ & $Q$ & $c_{VPP}$ \\\hline\hline
tree & $\rho$ & $-1/\sqrt{3}$ & - & - \\\hline
type-b & $\rho^-\pi^+$ & $-1/\sqrt{3}$ & $e$ & $\sqrt{2}$ \\
 & $\rho^+\pi^-$ & $-1/\sqrt{3}$ & $-e$ & $-\sqrt{2}$ \\
 & ${K^*}^-K^+$ & 1/2 & $e$ & $1/\sqrt{2}$ \\
 & ${K^*}^+K^-$ & 1/2 & $-e$ & $-1/\sqrt{2}$ \\\hline
type-c & $\rho^-\pi^+$ & $-1/\sqrt{3}$ & $-e$ & $\sqrt{2}$ \\
 & $\rho^+\pi^-$ & $-1/\sqrt{3}$ & $e$ & $-\sqrt{2}$ \\
 & ${K^*}^-K^+$ & 1/2 & $-e$ & $1/\sqrt{2}$ \\
 & ${K^*}^+K^-$ & 1/2 & $e$ & $-1/\sqrt{2}$ \\
\end{tabular}
\end{center}
\caption{Coefficients in eqs.~(\ref{eq:tree}),
(\ref{eq:gAVP}), (\ref{eq:typeb}) and
 (\ref{eq:typec}) for $h_1(1170/1380)\rightarrow \pi^0 \gamma$ decay.}
\label{tab:h1}
\end{table}

\begin{table}[h]
\begin{center}
\begin{tabular}{rc|r|r|r} 
\multicolumn{2}{c|}{$h_1(1170/1380)\rightarrow\eta\gamma$} & ${\cal C}$ & $Q$ & $c_{VPP}$ \\\hline\hline
tree & $\phi$ & 1 & - & - \\
 & $\omega$ & 1 & - & - \\\hline
type-b & ${K^*}^-K^+$ & 1/2 & $e$ & $\sqrt{3/2}$ \\
 & ${K^*}^+K^-$ & 1/2 & $-e$ & $-\sqrt{3/2}$ \\\hline
type-c  & ${K^*}^-K^+$ & 1/2 & $-e$ & $\sqrt{3/2}$ \\
 & ${K^*}^+K^-$ & 1/2 & $e$ & $-\sqrt{3/2}$ \\
\end{tabular}
\end{center}
\caption{Coefficients in eqs.~(\ref{eq:tree}),
(\ref{eq:gAVP}), (\ref{eq:typeb}) and
 (\ref{eq:typec}) for $h_1(1170/1380)\rightarrow \eta \gamma$ decay.}
\label{tab:h1eta}
\end{table}

\begin{table}[h]
\begin{center}
\begin{tabular}{rc|r|r|r} 
\multicolumn{2}{r|}{$b_1^0(1235)\rightarrow\pi^0\gamma$} & ${\cal C}$ & $Q$
 & $c_{VPP}$ \\\hline\hline 
tree & $\phi$ & 1 & - & - \\
 & $\omega$ & 1 & - & - \\\hline
type-b & ${K^*}^-K^+$ & $-1/2$ & $e$ & $1/\sqrt{2}$ \\
 & ${K^*}^+K^-$ & $-1/2$ & $-e$ & $-1/\sqrt{2}$ \\\hline
type-c & ${K^*}^-K^+$ & $-1/2$ & $-e$ & $1/\sqrt{2}$ \\
 & ${K^*}^+K^-$ & $-1/2$ & $e$ & $-1/\sqrt{2}$ \\
\end{tabular}
\end{center}
\caption{The coefficients for $b_1^0(1235)\rightarrow \pi^0 \gamma$ decay.}
\label{tab:b1}
\end{table}

\begin{table}[h]
\begin{center}
\begin{tabular}{rc|r|r|r} 
\multicolumn{2}{r|}{$b_1^0(1235)\rightarrow\eta\gamma$} & ${\cal C}$ & $Q$
 & $c_{VPP}$ \\\hline\hline 
tree & $\rho$ & 1 & - & - \\
\hline
type-b & ${K^*}^-K^+$ & $-1/2$ & $e$ & $\sqrt{3/2}$ \\
 & ${K^*}^+K^-$ & $-1/2$ & $-e$ & $-\sqrt{3/2}$ \\\hline
type-c & ${K^*}^-K^+$ & $-1/2$ & $-e$ & $\sqrt{3/2}$ \\
 & ${K^*}^+K^-$ & $-1/2$ & $e$ & $-\sqrt{3/2}$ \\
\end{tabular}
\end{center}
\caption{The coefficients for $b_1^0(1235)\rightarrow \eta \gamma$ decay.}
\label{tab:b1eta}
\end{table}

\begin{table}[h]
\begin{center}
\begin{tabular}{rc|r|r|r} 
\multicolumn{2}{c|}{$K_1^+(1270)\rightarrow K^+\gamma$}& ${\cal C}$ & $Q$ & $c_{VPP}$ \\
\hline\hline
tree & $\phi$ & 1 & - & - \\
 & $\omega$ & 1 & - & - \\
 & $\rho$ & $-1/\sqrt{3}$ & - & - \\
\hline
 type-b & $\phi K^+$ & 1 & $e$ & 1 \\
 & $\omega K^+$ & 1 & $e$ & $-1/\sqrt{2}$ \\
 & $\rho^0 K^+$ & $-1/\sqrt{3}$ & $e$ & $-1/\sqrt{2}$ \\
 & ${K^*}^0 \pi^+$ & $\sqrt{2/3}$ & $e$ & 1 \\
\hline
type-c & $\rho^+K^0$ & $-\sqrt{2/3}$ & $e$ & $-1$ \\
 & ${K^*}^+ \eta$ & 1 & $e$ & $\sqrt{3/2}$ \\
 & ${K^*}^+ \pi^0$ & $1/\sqrt{3}$ & $e$ & $1/\sqrt{2}$ \\
\end{tabular}
\end{center}
\caption{The coefficients in eqs.(\ref{eq:tree}), (\ref{eq:typeb}) and
 (\ref{eq:typec}) for ${K_1}^+\rightarrow K^+ \gamma$ decay.
}
\label{tab:Kplus}
\end{table}

\begin{table}[h]
\begin{center}
\begin{tabular}{rc|r|r|r} 
\multicolumn{2}{c|}{$K_1^0(1270)\rightarrow K^0\gamma$}& ${\cal C}$ & $Q$ & $c_{VPP}$ \\
\hline\hline
tree & $\phi$ & 1 & - & - \\
 & $\omega$ & 1 & - & - \\
 & $\rho$ & $1/\sqrt{3}$& - & - \\
\hline
type-b & $\rho^-K^+$ & $-\sqrt{2/3}$ & $e$ & $-1$ \\
 & ${K^*}^+\pi^-$ & $\sqrt{2/3}$& $-e$ & 1 \\
\hline
type-c & $\rho^-K^+$ & $-\sqrt{2/3}$  & $-e$ & $-1$ \\
 & ${K^*}^+\pi^-$ & $\sqrt{2/3}$& $e$ & 1 \\
\end{tabular}
\end{center}
\caption{The coefficients in eqs.(\ref{eq:tree}), (\ref{eq:typeb}) and
 (\ref{eq:typec}) for ${K_1}^0\rightarrow K^0 \gamma$ decay.
}
\label{tab:K0}
\end{table}

\noindent Eq.~(\ref{eq:tree}) is formally not gauge
invariant. An alternative derivation using tensor formalism is
given in ref.~\cite{Roca:2003uk}
and replaces $\epsilon_A\cdot\epsilon$ by
$(\epsilon_A\cdot\epsilon-\epsilon_A\cdot k\,\epsilon\cdot P/k\cdot P)$
with $P$ and $k$ the axial-vector meson and
photon momenta respectively.
Then the amplitude becomes
manifestly gauge invariant and reduces to Eq.~(\ref{eq:tree}) in
the Coulomb gauge ($\epsilon^0=0$)
which we use to evaluate the amplitudes.

In ref.~\cite{Roca:2006am}, the 
contribution to the total amplitude from 
type-b loops is shown to be
convergent by invoking gauge invariance.
This amplitude is given by
\begin{eqnarray}
t_{\rm b} &=& - g'_{AVP} Q c_{VPP} \frac{M_V G_V}{\sqrt{2}f^2} 2P\cdot k
\epsilon_A \cdot \epsilon \nonumber\\
&\times
&\int_0^1 dx \int_0^x dy
\frac{1}{32\pi^2}
\frac{1}{s+i\varepsilon}
\left\{-4(1-x)\left(1+\frac{y(xP-yk)\cdot(k-P)}{M_V^2}\right)\right\},
\label{eq:typeb}
\end{eqnarray}
where $Q$ is the charge of the meson in the loop emitting
 the photon and 
$c_{VPP}$ are numerical coefficients coming from the $VPP$
Lagrangian \cite{Roca:2006am} and are given in 
tables~\ref{tab:h1}--\ref{tab:K0}.
In Eq.~(\ref{eq:typeb}) $G_V$ is the $VPP$ coupling in the notation
of \cite{Ecker:1988te} and for the numerical value we take 
$G_V=55\pm 5$~MeV from ref.~\cite{Palomar:2003rb}, $f$ is the pion
decay constant ($f=93\mev$), $\epsilon_{A(\gamma)}$ is the
axial-vector(photon) polarization vector.
The coefficients shown in these tables for the channels with an
$\eta$ meson in the final state are for the decay into $\eta_8$.
Hence, in order to obtain the appropriate width of the channels
decaying into $\eta\gamma$
we have to multiply 
the decay width for these channels by $8/9$, from the
consideration of the mixing $\eta=\eta_1/3+2\sqrt{2}/3\eta_8$.

The amplitude from the type-c loop, neglecting the formally
logarithmically divergent but very small $1/M_V^2$ terms
 \cite{Roca:2006am}, is given by:
\begin{eqnarray}
t_{\rm c} &=& g'_{AVP} Q c_{VPP} \frac{M_VG_V}{\sqrt{2}f^2} 2P\cdot k
\epsilon_A\cdot\epsilon \nonumber\\
&\times&
\int_0^1 dx \int_0^x dy
\frac{1}{32\pi^2}\frac{1}{s'+i\varepsilon}
(1-3x+2y+y(1-x)).
\label{eq:typec}
\end{eqnarray}

With these amplitudes, the decay width
for the axial-vector mesons into one pseudoscalar
 meson and one photon is given by
\begin{equation}
\Gamma(M_A) = \frac{|\vec{k}|}{12\pi M_A^2} |T|^2,
\end{equation}
where $M_A$ stands for the mass of the decaying axial-vector meson
and $T$ is the sum of the amplitudes from the tree level and loop
mechanisms removing the $\epsilon_A\cdot\epsilon$ factor.
The former expression is valid in the limit of narrow axial-vector
resonance.
In order to take into account the finite width of the axial-vector
meson we fold the previous expression
with the mass distribution:
\begin{equation}
\Gamma_{A\rightarrow P\gamma} = -\frac{1}{\pi}
\int_{(M_A-2\Gamma_A)^2}^{(M_A+2\Gamma_A)^2}
ds_A\,
Im
\left\{\frac{1}{s_A-M_A^2+iM_A\Gamma_A}\right\}
\Gamma(\sqrt{s_A})
\Theta(\sqrt{s_A}-\sqrt{s_A^{th}}),
\end{equation}
where $\Theta$ is the step function, $\Gamma_A$ is the total
axial-vector meson width and $s_A^{th}$ is the threshold for the
dominant $A$ decay channels.

Similarly, since the $\rho$ and $K^*$ mesons have relatively large
widths, we have also taken into account the mass distribution of
these states in the loop functions leading to the $t_b$ and $t_c$
amplitudes. This is done by folding $t_b$, $t_c$, with the spectral
function of the $\rho$ and $K^*$:

\begin{equation}
t_{b,c}\to t_{b,c}= -\frac{1}{\pi}
\int_{(M_V-2\Gamma_V)^2}^{(M_V+2\Gamma_V)^2}
ds_V\,
Im
\left\{\frac{1}{s_V-M_V^2+iM_V\Gamma_V}\right\}
t_{b,c}(\sqrt{s_V}).
\end{equation}
\noindent
The corrections from this source are small, they change the
radiative widths at the level of $2$\% or below, although the
contribution of some intermediate states, which are particularly
suppressed, experiences larger
changes.

\section{Results for the different channels}

In what follows we discuss in detail the results for the radiative
decays of the different axial-vector resonances. We will refer to
the results shown in  tables~\ref{tab:width_h1}--\ref{tab:width_K0}
where we show the contributions of the different mechanisms 
to the radiative decay widths. The theoretical
errors quoted have been obtained by doing a Monte-Carlo sampling of
the parameters of the model within their uncertainties, as
explained in  ref.~\cite{Roca:2006am}. Note that in
ref.~\cite{Roca:2005nm} all the pole positions, and hence the
couplings to the different channels, were obtained with the same
value for the only free parameter of the model, the subtraction
constant, of $a=-1.85$ (see reference \cite{Roca:2005nm} for
details). But we can use different subtraction constants for
different ($S$, $I$, $G$-parity) channels. Therefore, in order to get
a more 
accurate result, we have fine tuned the subtraction
constants such that the real part of the pole positions agrees
better with the experimental axial-vector masses. On the other
hand, one can also assign an uncertainty to the $f$ constant 
appearing in the Lagrangians since it could range from $f_\pi$ to
$f_\eta$, averaging $1.15\times 92$~MeV. We have also considered
this uncertainty in our calculations. The central values in the
tables are obtained using $f=1.08 \times f_\pi$ and the central
values for the rest of the parameters (except for the $K_1$ case
where the more refined values of ref.~\cite{Geng:2006yb} for the
couplings and a different central value for $f$ are used).
 On the other hand, an extra conservative $10$\% 
has been added in quadrature to the error in order to consider the
uncertainty from the neglect of the
$1/M_V^2$ terms in the evaluation of the type-c loop contribution,
as explained in the previous section.


%


\begin{figure}[ht]
\epsfxsize=12cm
\centerline{
\epsfbox{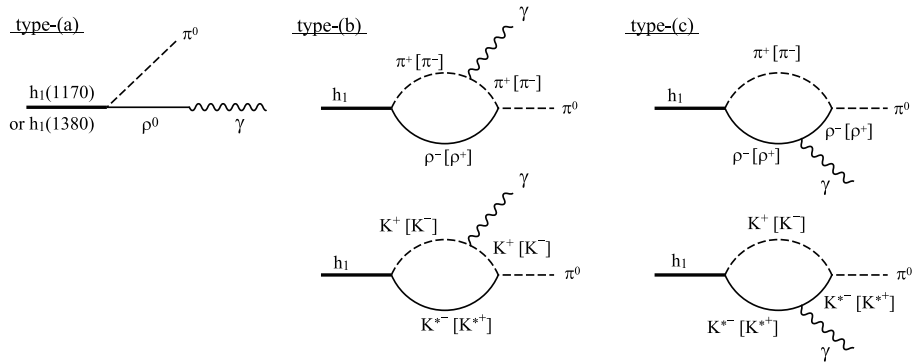}}
\caption{Feynman diagrams contributing to $h_1(1170)\rightarrow \pi^0\gamma$ decay and $h_1(1380)\rightarrow \pi^0\gamma$ decay.}
\label{fig:h1}
\end{figure}


\begin{figure}[ht]
\epsfxsize=12cm
\centerline{
\epsfbox{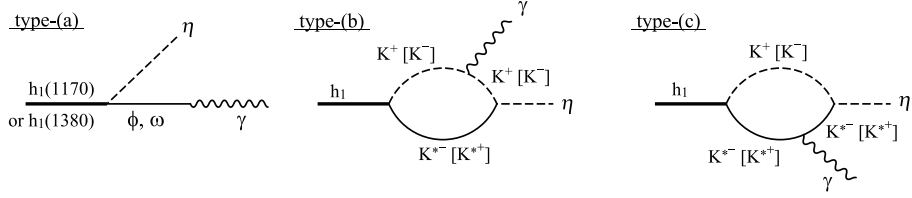}}
\caption{Feynman diagrams contributing to $h_1(1170)\rightarrow \eta\gamma$ decay and $h_1(1380)\rightarrow \eta\gamma$ decay.}
\label{fig:h1_eta}
\end{figure}


\begin{figure}[ht]
\epsfxsize=12cm
\centerline{
\epsfbox{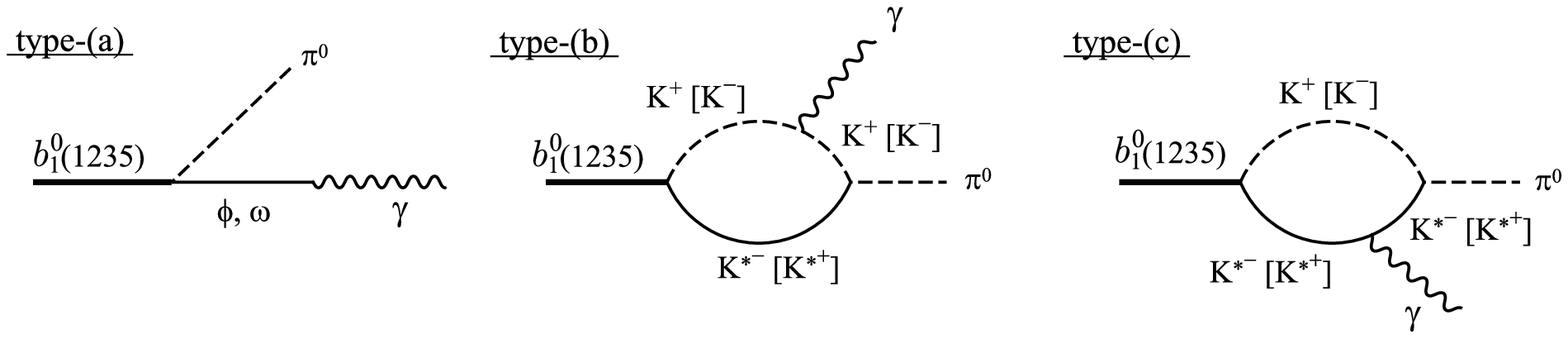}}
\caption{Feynman diagrams contributing to $b_1^0(1235)\rightarrow \pi^0\gamma$ decay.}
\label{fig:b1}
\end{figure}

\begin{figure}[ht]
\epsfxsize=12cm
\centerline{
\epsfbox{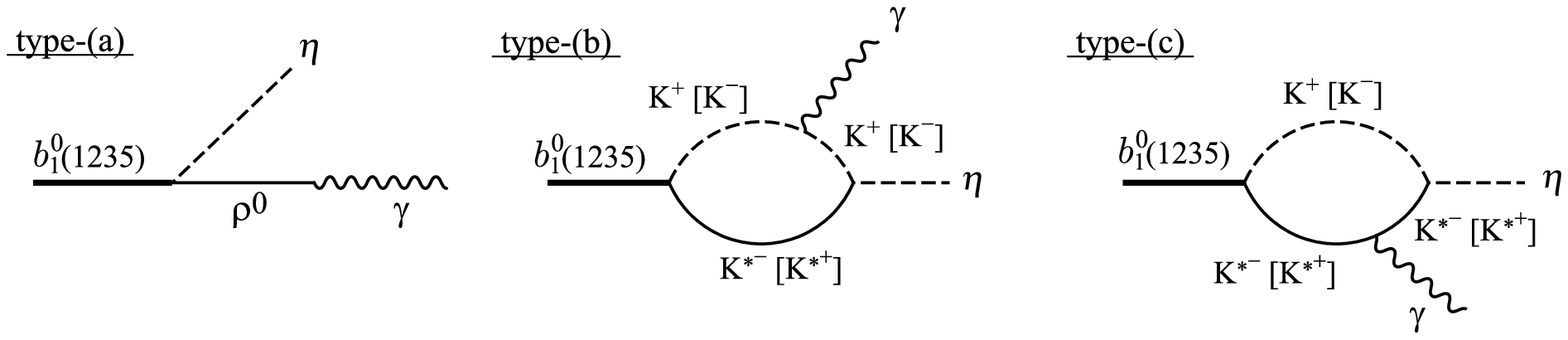}}
\caption{Feynman diagrams contributing to $b_1^0(1235)\rightarrow \eta\gamma$ decay.}
\label{fig:b1_eta}
\end{figure}


\begin{figure}[ht]
\epsfxsize=12cm
\centerline{
\epsfbox{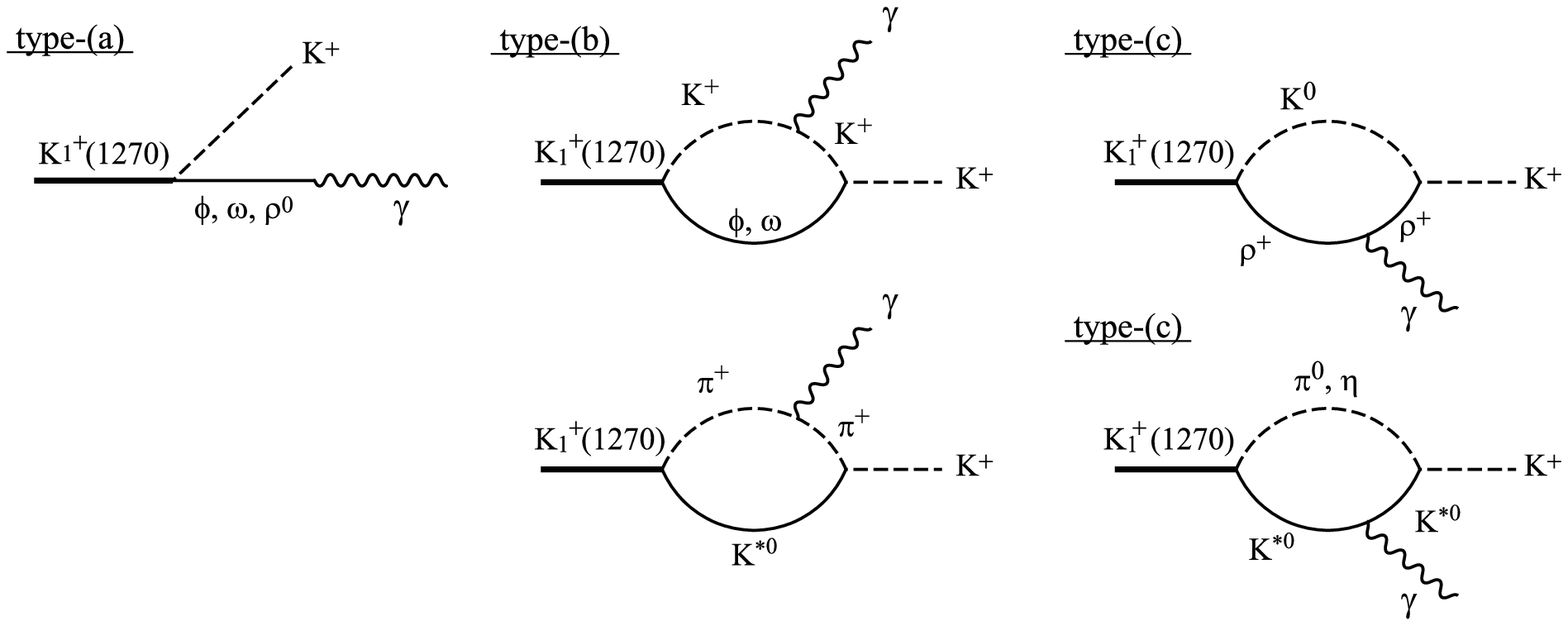}}
\caption{Feynman diagrams contributing to $K_1^+(1270)\rightarrow K^+\gamma$ decay.}
\label{fig:Kplus}
\end{figure}


\begin{figure}[ht]
\epsfxsize=12cm
\centerline{
\epsfbox{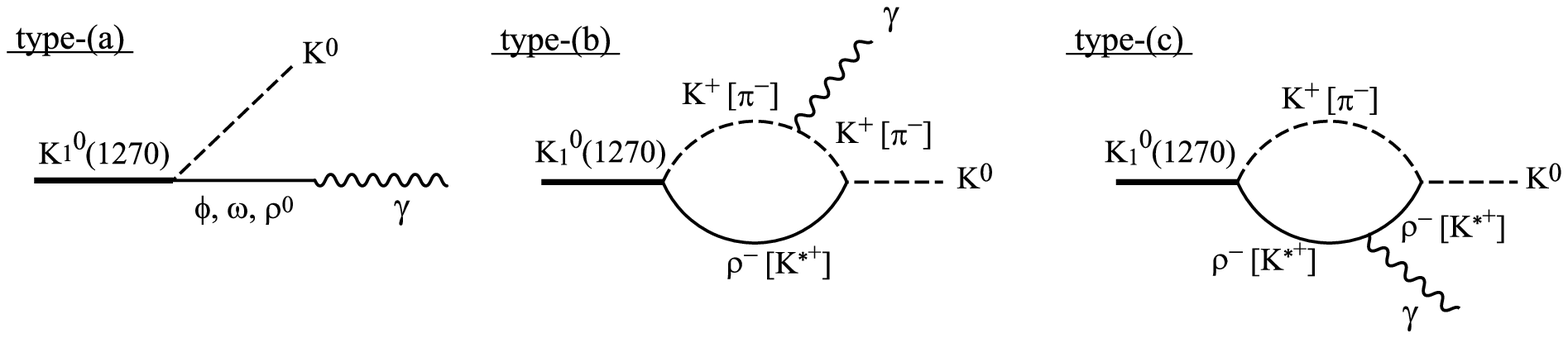}}
\caption{Feynman diagrams contributing to $K_1^0(1270)\rightarrow K^0 \gamma$ decay.}
\label{fig:K0}
\end{figure}


\begin{table}[ht]
\begin{center}
\begin{tabular}{cc|r|r} 
 &  & \multicolumn{1}{c|}{$h_1(1170)$} & \multicolumn{1}{c}{$h_1(1380)$} \\
\hline\hline
tree & $\rho$ & 294.8 & 9.1 \\
\cline{2-4}
 & total & 294.8 & 9.1 \\
\hline\hline
type-b & $\rho^- \pi^+$ & 51.8 & 2.2 \\
 & $\rho^+ \pi^-$ & \multicolumn{2}{c}{same as $\rho^- \pi^+$} \\
 & ${K^*}^- K^+$ & 0.43 & 15.2 \\
 & ${K^*}^+ K^-$ & \multicolumn{2}{c}{same as ${K^*}^- K^+$} \\
\cline{2-4}
 & total & 226.0 & 37.0 \\
\hline
type-c & $\rho^- \pi^+$ & 0.30 & $1.5\times 10^{-2}$ \\
 & $\rho^+ \pi^-$ & \multicolumn{2}{c}{same as $\rho^- \pi^+$} \\
 & ${K^*}^- K^+$ & $2.2\times 10^{-3}$ & $6.5\times 10^{-2}$ \\
 & ${K^*}^+ K^-$ & \multicolumn{2}{c}{same as ${K^*}^- K^+$} \\
\cline{2-4}
 & total & 1.34 & 0.51 \\
\hline
 & loop total & 218.0 & 45.8 \\
\hline\hline
\multicolumn{2}{c|}{TOTAL} & $837\pm 134$ & $81\pm 18 $ \\
\end{tabular}
\caption{$h_1(1170)$/$h_1(1380)\rightarrow \pi^0 \gamma$ decay widths in
 units of keV.}
\label{tab:width_h1}
\end{center}
\end{table}

\begin{table}[ht]
\begin{center}
\begin{tabular}{cc|r|r} 
 &  & \multicolumn{1}{c|}{$h_1(1170)$} & \multicolumn{1}{c}{$h_1(1380)$} \\
\hline\hline
tree & $\phi$ & $2.0\times 10^{-2}$ & 36.3 \\
   & $\omega$ & $3.5\times 10^{-3}$ & 21.3 \\
\cline{2-4}
 & total & $1.24\times 10^{-2}$ & 113.3 \\
\hline\hline
type-b & ${K^*}^- K^+$ & 0.75 & 28.5 \\
 & ${K^*}^+ K^-$ & \multicolumn{2}{c}{same as ${K^*}^- K^+$} \\
\cline{2-4}
 & total & 3.01 & 113.9 \\
\hline
type-c & ${K^*}^- K^+$ & $2.71\times 10^{-3}$ & 0.085 \\
 & ${K^*}^+ K^-$ & \multicolumn{2}{c}{same as ${K^*}^- K^+$} \\
\cline{2-4}
 & total & $1.08\times 10^{-2}$ & 0.34 \\
\hline
 & loop total & 3.37 & 126.6 \\
\hline\hline
\multicolumn{2}{c|}{TOTAL} & $3.1 \pm 0.9 $ & $438\pm 80 $ \\
\end{tabular}
\caption{$h_1(1170)$/$h_1(1380)\rightarrow \eta \gamma$ decay widths in
 units of keV.}
\label{tab:width_h1eta}
\end{center}
\end{table}

\begin{table}
\begin{center}
\begin{tabular}{cc|r} 
 &  & $b_1^0(1235)$ \\
\hline
tree & $\phi$ & 19.7 \\
 & $\omega$ & 14.1 \\
\cline{2-3}
 & total & 66.9 \\
\hline\hline
type-b & ${K^*}^- K^+$ & 6.6 \\
 & ${K^*}^+ K^-$ & same as ${K^*}^- K^+$ \\
\hline
 & total & 26.3 \\
\hline\hline
type-c & ${K^*}^- K^+$ & $3.3\times 10^{-2}$ \\
 & ${K^*}^+ K^-$ & same as ${K^*}^- K^+$ \\
\cline{2-3}
 & total & 0.13 \\
\hline
 & loop total & 30.1 \\
\hline\hline
\multicolumn{2}{c|}{TOTAL} & $180\pm 28$ \\
\end{tabular}
\caption{$b_1^0(1235)\rightarrow \pi^0 \gamma$ decay width in
 units of keV.}
\label{tab:width_b1}
\end{center}
\end{table}

\begin{table}
\begin{center}
\begin{tabular}{cc|r} 
 &  & $b_1^0(1235)$ \\
\hline
tree & $\rho$ & 244.0 \\
\cline{2-3}
 & total & 244.0 \\
\hline\hline
type-b & ${K^*}^- K^+$ & 10.8 \\
 & ${K^*}^+ K^-$ & same as ${K^*}^- K^+$ \\
\hline
 & total & 43.2 \\
\hline\hline
type-c & ${K^*}^- K^+$ & $3.8\times 10^{-2}$ \\
 & ${K^*}^+ K^-$ & same as ${K^*}^- K^+$ \\
\cline{2-3}
 & total & 0.15 \\
\hline
 & loop total & 48.5 \\
\hline\hline
\multicolumn{2}{c|}{TOTAL} & $488\pm 70$ \\
\end{tabular}
\caption{$b_1^0(1235)\rightarrow \eta \gamma$ decay width in
 units of keV.}
\label{tab:width_b1eta}
\end{center}
\end{table}

\begin{table}
\begin{center}
\begin{tabular}{cc|rr} 
&&\multicolumn{2}{c}{$K_1(1270)$}\\ \hline
&&pole-A & pole-B\\
 &  & $1195-123i$ & $1284-73i$ \\ \cline{3-4}
\hline\hline
tree & $\phi$ & 24.7 & 8.3 \\
 & $\omega$ & 17.4 &  5.3 \\
 & $\rho^0$ & 58.4 & 253.8 \\
\cline{2-4}
 & total & 8.58 & 412.5 \\
\hline\hline
type-b & $\phi K^+$ & 2.4 & 0.80 \\
 & $\omega K^+$ & 3.3 & 1.0 \\
 & $\rho^0 K^+$ & 0.9 & 3.9 \\
 & ${K^*}^0 \pi^+$ & 41.9 & 2.7 \\
\cline{2-4}
 & total & 68.4 & 25.7 \\
\hline
type-c & $\rho^+ K^0$ & $3.0\times 10^{-2}$ & 0.13 \\
 & ${K^*}^+ \eta$ & $2.3\times 10^{-4}$ & $6.4\times 10^{-2}$ \\
 & ${K^*}^+ \pi^0$ & 0.12 & $7.5\times 10^{-3}$ \\
\cline{2-4}
 & total & 0.15 & 0.26 \\
\hline
 & loop total & 75.3 & 20.9 \\
\hline\hline
\multicolumn{2}{c|}{TOTAL} & $34\pm 13$ & $251\pm 56$ \\
\end{tabular}
\caption{$K_1^+(1270) \rightarrow K^+ \gamma$ widths for two poles in
 units of keV.}
\label{tab:width_Kplus}
\end{center}
\end{table}

\begin{table}
\begin{center}
\begin{tabular}[h]{cc|rr}
&&\multicolumn{2}{c}{$K_1(1270)$}\\ \hline
&&pole-A & pole-B\\
&& $1195-123i$ & $1284-73i$ \\
\hline\hline
tree&$\phi$ & 24.7 & 8.3 \\
&$\omega$ & 17.4 & 5.3 \\
&$\rho_0$ & 58.4 & 253.8\\
\cline{2-4}
&total & 274.3 & 148.7 \\
\hline\hline
type-b &$\rho^-K^+$ & 3.6 & 15.5\\
&${K^*}^+\pi^-$ & 41.8&2.7\\
\cline{2-4}
&total&61.5&6.7\\
\hline
type-c&$\rho^-K^+$ & $3.0 \times 10^{-2}$ & 0.13\\
&${K^*}^+\pi^-$ & 0.47& $3.0\times 10^{-2}$\\
\cline{2-4}
&total&0.48&0.28\\
\hline
& loop total & 57.6 & 9.7 \\
\hline\hline
\multicolumn{2}{c|}{TOTAL} & $512\pm 73$ & $227 \pm 79 $\\
\end{tabular}
\caption{$K_1^0(1270) \rightarrow K^0 \gamma$ widths for two poles
in units of keV.}
\label{tab:width_K0}
\end{center}
\end{table}


\subsection{$S=0$, $I=0$ channel}
\subsubsection*{$h_1(1170)$/$h_1(1380)\rightarrow \pi^0 \gamma$}

The $S=0$, $I=0$ and negative $G$-parity axial-vector mesons
couple to $\phi\eta$, $\omega\eta$, $\rho\pi$ and the combination
$1/\sqrt{2}(\bar{K^*}K-K^*\bar{K})$ in our model.
 However, the first two channels lead to b and c type diagrams with
 neutral intermediate mesons and do not contribute.
 Hence, only the diagrams shown in fig.~\ref{fig:h1}, with 
$\rho\pi$ and $K^*K$ in the loops,
contribute to the process.
For the tree level diagram only the $\rho$ meson exchange 
is possible. 
In our model, the $h_1(1170)$ resonance
 has a coupling to the
$\rho\pi$ channel that is about five times 
the one of $h_1(1380)$.
Altogether this makes the tree level and the $\rho\pi$ loop 
contributions much larger for  the $h_1(1170)$ decay
than for the $h_1(1380)$ one, as seen in 
table~\ref{tab:width_h1}.
This implies at the end, after the coherent sum of all the
contributions, that the radiative decay 
width of the $h_1(1170)$ into
$\pi\gamma$ is much larger than that of
 the $h_1(1380)$ and, hopefully,
it could be measured experimentally given its large
 value $863\pm 134$~keV.
Note the important role of the loop contribution which makes that
the final result obtained for the $h_1(1170)$  
is about a factor $3$ larger than
considering only the tree level mechanism and about a factor $9$
for the $h_1(1380)$ case.

\subsubsection*{$h_1(1170)$/$h_1(1380)\rightarrow \eta \gamma$}
The diagrams needed in the evaluation of this process 
are shown in fig.\ref{fig:h1_eta}. The couplings of the $h_1(1170)$ to the
$\omega\eta$ and $\phi\eta $ channels are very small. This makes
the tree level almost negligible. This is not the case for the
$h_1(1380)$ resonance since its couplings to these channels
are much larger.
 On the other hand, the loop contributions are also smaller for the
$h_1(1170)$ since its coupling
to $1/\sqrt{2}(\bar{K^*}K-K^*\bar{K})$ is about
a factor four smaller than that of the $h_1(1380)$ resonance. 
All these facts, together with the constructive interference
between the loops and the tree level in the $h_1(1380)$ case,
 make the radiative decay width of the $h_1(1380)$ two
orders of magnitude larger than the $h_1(1170)$.

\subsubsection*{$f_1(1285) \rightarrow \pi^0 \gamma$}

This channel is zero by ${\cal C}$-parity conservation.

\subsection{$S=0$, $I=1$ channel}

The radiative decays into $\pi\gamma$ of the charged
$a_1(1260)$ and $b_1(1235)$  resonances 
were thoroughly discussed in
ref.~\cite{Roca:2006am},  and
hence we only consider in the present paper 
 the neutral $S=0$, $I=1$, axial-vector
radiative decay modes. 

\subsubsection*{$a_1^0(1260)\rightarrow \pi^0\gamma$}

This channel is zero by ${\cal C}$-parity conservation.
However, as seen in ref.~\cite{Roca:2006am},
the charged decay channel,
 $a_1^\pm(1260)\rightarrow \pi^\pm\gamma$, was allowed and had a
 large decay width.

\subsubsection*{$b_1^0(1235)\rightarrow \pi^0\gamma$}

The $b_1(1235)$ couples to the positive $G$-parity $VP$ states 
$1/\sqrt{2}(\bar{K^*}K+K^*\bar{K})$, $\phi\pi$, $\omega\pi$ and
$\rho\eta$.
The allowed Feynman diagrams are shown in fig.~\ref{fig:b1}. 
One can see from table~\ref{tab:width_b1}
 that the tree level 
only accounts for
$\sim1/3$ of the final result. This illustrates the important role
of the loops considered in the present formalism.
It is worth stressing that the result obtained for the
 $b_1^0(1235)\rightarrow \pi^0\gamma$ decay is the 
 same\footnote{The numerical difference with the result in 
ref.~\cite{Roca:2006am} 
is the different central value used for $f$ and
 hence the different central values for the couplings, as
 explained above. The differences are within the theoretical
 uncertainties estimated in each case.}
 as for the 
$b_1^\pm(1235)\rightarrow \pi^\pm\gamma$ decay obtained in 
\cite{Roca:2006am}, unlike the $a_1(1260)$
case, as explained above.

\subsubsection*{$b_1^0(1235)\rightarrow \eta\gamma$}
The allowed Feynman diagrams are shown in fig.~\ref{fig:b1_eta}.
As can be seen in table~\ref{tab:width_b1eta}, 
the tree level contribution to the decay width 
for this channel is much larger than 
in the $b_1^0(1235)\rightarrow \pi^0\gamma$ (despite the
phase space available) since the coupling to
$\rho\eta$ is larger than to $\omega\pi$ and $\phi\pi$.
This size of the tree level, together with the constructive
interference with the loop contributions, makes the final radiative
 width for this channel very large, about a factor $\sim 3$
 larger than the $b_1^0(1235)\rightarrow \pi^0\gamma$ decay rate.

\subsection{Consideration of higher mass intermediate states}

In principle we could consider in our approach the contribution of
additional channels involving vector-pseudoscalar states of higher
masses. The contributions of such channels in the chiral unitary
approach leading to the axial-vector mesons \cite{Roca:2005nm} was
omitted, as usually done in this approach. The idea is that they,
being far off shell in the loops, provide a small contribution, but
more important, they can be reabsorbed into the subtraction
constant of the dispersion relation which provides the loop
function, because their contribution is very weakly energy dependent.
In the present work, where the loops are finite, the contribution
from these channels would be additive. We make here an estimation
of the contribution of these heavy states.

The first consideration is that when one has many channels in the
chiral unitary approach, the coupling of the resonance to the
channels with mass far from the resonance mass is very weak as a
general rule. For instance, the coupling of the $\sigma(600)$ to
$\eta\eta$ is about $4$\% of the one to $\pi\pi$ 
\cite{Oller:1998zr}. The coupling of the $\Lambda(1405)$ to $K\Xi$
is of the order of one fourth of the dominant $\pi\Sigma$ and $\bar
K N$ channels. We should keep this fact in mind. To make the
estimates of the high mass states contribution in the present
case, we take intermediate states where the $\pi$ is replaced by
the $\pi(1300)$ and other states where the $\rho$ is substituted by
the $\rho(1450)$, the next pseudoscalar and vector excited states.
We assume first the loops changing $\pi$ to $\pi(1300)$. We take in
a first run the couplings in the loop the same as for $\pi$ and
investigate only the effect of the change in the mass. We get
contributions of the order of $2$\% of the contribution of the
loops of $\rho\pi$ in the $h_1$ decays and in the $K^*\pi$ of the
$K_1$ decay from the type-b loops. In type-c loops, for some
intermediate states the relative contribution of the $\pi(1300)$ with
respect to the $\pi$
case is larger but these terms have very small weight. Next we
should consider the difference between the $\rho\pi\pi$ and
$\rho\pi\pi(1300)$ couplings. By looking at $\rho\to\pi\pi$ and
$\pi(1300)\to\rho\pi$ decays, assuming $\Gamma(1300)\sim 400\mev$
from the PDG \cite{pdg} and all strength of $\pi(1300)$ going to
$\rho\pi$, we obtain a ratio of couplings
$g_{\pi\pi'\rho}/g_{\pi\pi\rho}=1.8$, which should be compensated
from the smaller coupling of the axial-vector resonances to the
$\rho\pi(1300)$ state. The effect at the end in the total radiative
width of the resonance is smaller than the one quoted above from
some particular channels and is far smaller than the uncertainties
in the results from the other sources considered here.

Next we do the same exercise by changing the $\rho$ to the
$\rho(1450)$. The change of the mass without changing coupling
constants is in general very small with the exception of the
contribution  of the intermediate state $\rho(1450)\pi$ for the two
$h_1$ decays. In this case the contribution of the new loop to
$t_b$ is $17$\% of the contribution of the $\rho(770)\pi$.
However, we should now take into account the ratio
$g_{\rho(1450)\pi\pi}/g_{\rho\pi\pi}=0.5$ following the same steps
as before, assuming in the worse of the cases  that all the
$\rho(1450)$ width comes from $\pi\pi$.
This changes in a maximum of $8$\% the contributions to $t_b$ from
this intermediate state. With a conservative estimate of a factor
of two reduction from the coupling of the axial-vector resonance to
this new channel, this results in a $4$\% change of the
contribution to $t_b$ from the $\rho\pi$ channel. When this new
contribution is added to all other terms it has a repercussion of a
maximum $4$\% change in the total decay rate for the $h_1(1170)$
and much smaller for the $h_1(1380)$. Once again this
uncertainty is smaller than the one obtained before from other
sources.

\section{Consequences of the two $K_1(1270)$ poles}

We have singled out the $S=1$, $I=1/2$, sector into this different
section since it deserves particular attention for the following
reasons. In the first place, in the work of 
ref.~\cite{Roca:2005nm}, two poles were found in the
 $S=1$, $I=1/2$, $VP$ scattering amplitude
which were assigned there to two $K_1(1270)$ resonances
instead of the usual $K_1(1270)$ and $K_1(1400)$.
 In
table~\ref{tab:g} we show the two pole positions
and the couplings to the different $VP$ channels of the two
$K_1(1270)$ resonances obtained in \cite{Geng:2006yb}.
In the following, we call
pole-A the lowest mass pole
and pole-B the highest mass one.  Some possible experimental
consequences of this double pole structure of the $K_1(1270)$ 
resonance were
already discussed in  ref.~\cite{Geng:2006yb}. If this double pole
reflects the real nature of these resonance, it would have
significant relevance in the study of the radiative decays both
from the theoretical  and experimental points of view. Indeed,
there is experimental information \cite{Alavi-Harati:2001ic} on the
decay width of the process $K_1^0(1270) \rightarrow K^0 \gamma$ 
which relies in an experimental analysis that does
 not consider the two pole structure.
The result for the radiative width would change had the two
pole nature  of the $K_1(1270)$ been considered, as we will discuss
in this section.

First of all let us present our theoretical results for the
radiative decay widths of $K_1^\pm$ and $K^0$ for both poles A and
B. 

\subsubsection*{$K_1^+(1270)\rightarrow K^+ \gamma$}

The possible intermediate channels in the tree level and type-b and
-c loops for the $K^+_1(1270)$ radiative decay are shown in 
fig.~\ref{fig:Kplus}. The results for the different contributions
for both  poles A and B are shown 
in table~\ref{tab:width_Kplus}. In
this table one can see that the tree level contribution for the
pole-B is about a factor fifty  larger than for the pole-A. This is
due to the fact that the coupling of the pole-B to the  $\rho K$
channel is about a factor two larger than for the pole-A (see
table~\ref{tab:g}) and to the different sign of the $\rho K$ 
couplings for
both poles (see table~\ref{tab:g}), what makes  the
interference with the $\phi$ and $\omega$ contributions different.
 The
difference in the couplings 
to the $VP$ channels for the
pole-A and B, both in sign and absolute value, is also  responsible
for the different value of the loop contributions and the sign of
the interference, leading to a final result
for the radiative width which is about an order of magnitude 
larger for the
B-pole.

\subsubsection*{$K_1^0(1270) \rightarrow K^0 \gamma$}
 For the $K_1^0(1270)$ radiative decay,
the allowed mechanisms are shown in fig.~\ref{fig:K0}.
Note the different 
allowed particles
 in the loops with respect the $K_1^+(1270)$ case
  (fig.~\ref{fig:Kplus}),
like for instance the presence of $\rho$ instead of $\phi$ and
$\omega$ in the type-b loop.

Unlike the $K_1^+(1270)$ case,
we obtain similar values of
the tree level contributions for both poles, 
while the individual contributions to the tree level
are the same as in the
 $K_1^+$ case (see tables~\ref{tab:width_Kplus} and
\ref{tab:width_K0}). 
This is due to the different sign of $K_1^0$
to the $\rho K$ channel in the charge base from that of $K_1^+$,
 as
can be seen in tables~\ref{tab:Kplus} and \ref{tab:K0}.
Note that the ${K^*}\pi$ loop contribution
for the pole-A is larger
than the $\rho K$, while for the pole-B the $\rho K$
contribution is larger than $K^*\pi$.
This is a consequence of the fact that the largest coupling for the
pole-A is to
${K^*}\pi$ while for the pole-B is to the $\rho K$ channel. 

\subsection{Discussion on the experimental result}

Unlike the other channels, there is experimental information 
 about the $K_1^0(1270) \rightarrow K^0 \gamma$ decay width.
This radiative decay width has already been measured at
Fermilab~\cite{Alavi-Harati:2001ic}. 
In the analysis of this experiment \cite{Alavi-Harati:2001ic}
it was  concluded that the radiative
decay width of $K_1(1270)$ is $73.2\pm6.1 ({\rm stat})\pm 8.2({\rm int
\ syst})\pm27.0({\rm ext\ syst})$~keV and that of $K_1(1400)$ is 
$280.8 \pm 23.2({\rm stat})\pm31.4({\rm int\ stat})\pm25.4({\rm ext\
stat})$~keV. This is in remarkable disagreement with our
results mentioned above and in 
table~\ref{tab:width_K0}.
 However this discrepancy can
be explained in view of the two pole structure of the $K_1(1270)$
resonance, not considered in the experimental analysis:  

The experiment~\cite{Alavi-Harati:2001ic}
 did not measure the $K_1(1270)$ radiative decay
directly. 
Since direct observation of radiative decays such as
$K_1 \rightarrow K + \gamma$ is difficult, they measured the $K_1$
radiative decay using the inverse reaction $K + \gamma \rightarrow K_1$,
which can be performed experimentally using the $K+{\rm nucleus}
\rightarrow K_1 + {\rm nucleus}$ reaction with Primakoff
effect~\cite{Primakoff}. 
The experiment obtained 147 events for 
strange axial-vector mesons  
reconstructed from $K^*\pi$ final state
and used them to estimate the radiative
widths for $K_1(1270)$ and $K_1(1400)$.
But in the experimental analysis it is assumed that there is only 
one
$K_1(1270)$ and a $K_1(1400)$ resonances contributing to the
events.
The traditional approach to the $K_1(1270)$ and $K_1(1400)$
resonances, and the one assumed in the experiment,
 is that they are
     a mixture of a singlet and triplet component
\begin{eqnarray}
K_1(1400) &=& ^3P_1 \cos\theta + ^1P_1\sin\theta\\
K_1(1270) &=& -^3P_1 \sin\theta + ^1P_1\cos\theta.
\end{eqnarray}
 The value for the mixing angle used in the experimental analysis
 \cite{Alavi-Harati:2001ic} 
  is $\theta= 56^\circ \pm
3^\circ$ \cite{Daum:1981hb} but there is
 controversy about this value (see the
discussion in the introduction of ref.~\cite{Roca:2003uk})  and
very different mixing angles
 are quoted from the study of $J/\Psi$ decay
 \cite{Li:2006we}.
  As quoted in~\cite{Geng:2006yb} this  could be a problem
related to the existence of two $K_1(1270)$ resonances. Coming back
to the experimental analysis of ref.~\cite{Alavi-Harati:2001ic},
 since the triplet component is not
excited by the Coulomb field~\cite{Babcock:1976hr}, the $K_1$
production rates would be proportional to $\cos^2\theta$ for the
$K_1(1400)$ and $\sin^2 \theta$  for the $K_1(1270)$ from where the
ratio of  $72.2$~keV for the $K_1(1270)$ and $280.8$~keV  for the
$K_1(1400)$ was deduced in the experimental analysis. However,
 in our approach this mixing scheme would be different
since the mixing would be between two $K_1(1270)$ resonances, and
possibly a $K_1(1400)$, rather
than between only one $K_1(1270)$ and a $K_1(1400)$.
If there are actually two $K_1(1270)$ poles, instead of just one, 
this invalidates the different weights assigned to the
 $K_1(1270)$ and the $K_1(1400)$ in 
 ref.~\cite{Alavi-Harati:2001ic}. 

\begin{figure}[ht]
\epsfxsize=6cm
\centerline{
\epsfbox{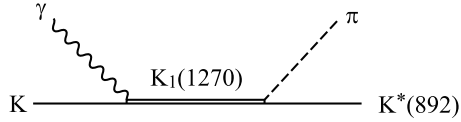}}
\caption{Subprocess in the experimental mechanism producing
the $K_1$ with $K^*\pi$ in the final state. }
\label{fig:primakoff}
\end{figure}

Let us hence make an alternative
guess. 
The experiment\cite{Alavi-Harati:2001ic}
 observed the ${K^*}^0(892)\pi^0$ channel in the final 
state, hence as a subprocess of the experimental reaction we have 
the mechanisms shown in fig.~\ref{fig:primakoff}.

In our model, we have two poles for the $K_1(1270)$ resonance.
In this case, the contributions of these two poles should
interfere. 
We can estimate the effect of this interference in the 
experiment as follows.
The amplitude of the subreaction process in the experiment, shown 
in fig.~\ref{fig:primakoff}, for the two different poles can be
 written approximately as
\begin{eqnarray} 
T_A &=& g_{{K^*}^0\pi^0}^Ag_{K^0\gamma}^A
 \frac{1}{{s_{K_1}-M^{A}_{K_1}}^2+iM^{A}_{K_1}
 \Gamma^{A}_{K_1}}, \nonumber\\
T_B &=& g_{{K^*}^0\pi^0}^Bg_{K^0\gamma}^B
 \frac{1}{{s_{K_1}-M^{B}_{K_1}}^2+iM^{B}_{K_1}
 \Gamma^{B}_{K_1}},
\label{eq:T2}
\end{eqnarray}
where $g_{{K^*}^0\pi^0}^{A(B)}$ is the coupling of the
$K_1^0(1270)$ pole-A(B) to ${K^*}^0\pi^0$ channel
in charge base as
$$
g_{{K^*}^0\pi^0}^{A(B)} = -\frac{1}{\sqrt{3}}g_{K^*\pi}^{A(B)}
$$
with $g_{K^*\pi}^{A(B)}$ being the coupling
of the pole-A(B) to $K^*\pi$ in isospin base.
In Eq.~(\ref{eq:T2}) $g_{K^0\gamma}^{A (B)}$ is the coupling constant
 of the $K_1^0(1270)$ pole-A(B)
to the $K^0\gamma$ which can be evaluated from the radiative 
decay amplitudes obtained in this paper
at $\sqrt{s_{K_1}}=M_{K_1}$ if we define 
$t_{K_1^0\to K^0\gamma}
=g_{K_1^0 K^0\gamma}\epsilon_A\cdot\epsilon$.
In Eq.~(\ref{eq:T2}) the masses and widths used in the
 propagators are the ones given by the poles of table~\ref{tab:g}. 
The couplings that we found for the different poles
 are shown in table~\ref{tab:g}.

\begin{table}[h]
\begin{center}
\begin{tabular}{c|cc|cc} 
\multicolumn{1}{r|}{} & \multicolumn{2}{c}{pole-A} & \multicolumn{2}{|c}{pole-B} \\\hline\hline
 & $g^A$ & $|g^A|$ & $g^B$ & $|g^B|$ \\
\hline
$K^0\gamma $&$ 217-96i $&$ 237 $&$ -166-66i $&$ 179 $\\
\hline
${K^*}^0\pi^0 $&$ -2740+1659i $&$ 3203 $&$ -444+676i $&$ 809 $\\
$\rho^0 K^0 $&$ -965+923i $&$ 1335 $&$ 2774+228i $&$ 2783 $\\
\end{tabular}
\end{center}
\caption{The coupling constants of the two
 $K_1^0(1270)$ to $K^0\gamma$, ${K^*}^0\pi^0$
 and $\rho^0 K^0$ in charge base.
}
\label{tab:g}
\end{table}

\begin{figure}[h]
\epsfxsize=13cm
\centerline{
\epsfbox{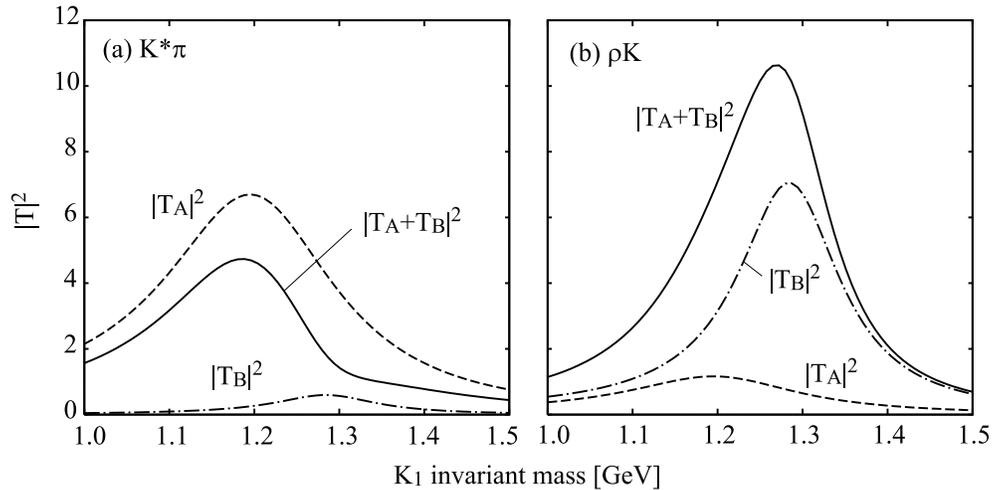}}
\caption{The squared amplitudes corresponding eqs.~(\ref{eq:T2}) and their
 coherent sums for (a) $K^*\pi$ and
(b) $\rho K$ in final states, as functions of the $K_1$ invariant mass.}
\label{fig:T2}
\end{figure}

As we can see from table~\ref{tab:g}, the coupling of the 
pole-A to the $K^*\pi$
channel is about four times larger  than the pole-B.
Hence, this process is dominated by the pole-A contribution.
In fig.~\ref{fig:T2}(a) we show the modulus squared of 
$T_A$, $T_B$, of Eq.~(\ref{eq:T2}) and
the coherent addition of both amplitudes.
Since the sign of the two $K_1$ couplings to
$K\gamma$ are different for both poles,
the interference between $T_A$ and
$T_B$ is destructive for the $K^*\pi$ case.
  This means that the amplitude 
observed in such a experiment
should be smaller than the one obtained with the dominant 
amplitude. The ratio
$|T_A+T_B|^2/|T_A|^2$ is $\sim 0.7$,
 and hence the corresponding radiative
decay width observed in such experiment from our model would be
$525$~keV$\times 0.7 = 370$~keV. 
This value is much similar to the addition of
the experimental values of the radiative decay width of the 
$K_1(1270)$ and $K_1(1400)$ of ref.~\cite{Alavi-Harati:2001ic},
 $353\pm 55$~keV.
 In other words, the experiment sees
the addition of the decay widths of the different $K_1$ resonances.
Of course, the peak seen in ref.~\cite{Alavi-Harati:2001ic}
seems to have an appreciable contribution from the $K_1(1400)$
resonance and a model independent way of separating it would be
most welcome. Note however that because the $K_1(1400)$ shares the
same quantum numbers as the $K_1(1270)$ one should sum coherently
the resonant amplitudes instead of assuming an incoherent sum of
decay rates as done in ref.~\cite{Alavi-Harati:2001ic}.
On the other hand, the analysis  
in ref.~\cite{Alavi-Harati:2001ic} relies on a coupling of the
$K_1(1270)$ resonance to $K^*\pi$ extracted from the information of
the PDG \cite{pdg} which is also questioned in ref.~\cite{Geng:2006yb} in base
of the existence of the two $K_1(1270)$ resonances, which have very
different coupling to $K^*\pi$.
Certainly, the study of a reaction where the $K_1(1400)$ production
would be suppressed would be a most suitable reaction to measure the
$K_1(1270)$ properties. In the next subsection we address such a
reaction.

\subsection{Primakoff reaction with $\rho K$ final state}

\begin{figure}[h]
\epsfxsize=6cm
\centerline{
\epsfbox{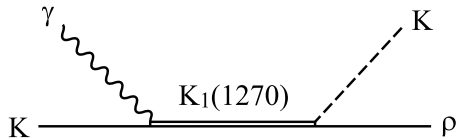}}
\caption{The same process as in fig.~\ref{fig:primakoff} but
 $\rho K$ in final state.}
\label{fig:primakoff_rhoK}
\end{figure}
 
The same reaction as in ref.~\cite{Alavi-Harati:2001ic} but looking
at $\rho K$ in the final state would have the advantage that the
$K_1(1400)$ has a negligible decay rate to $\rho K$ \cite{pdg}.
Hence, the $K_1(1270)$ resonance would stand clean in the
reaction.  A similar  analysis as done in
ref.~\cite{Alavi-Harati:2001ic} could now be done in order to
obtain the $K_1(1270)\to K^0\gamma$ decay width and
 would serve as a
test of consistency of the result obtained in 
ref.~\cite{Alavi-Harati:2001ic}. However, from the perspective of
the two pole scenario, this consistency 
is unlikely as we show below. Indeed,
in fig.~\ref{fig:T2}(b) we show the analogous to  \ref{fig:T2}(a)
for the $\rho K$ case. As we can see in the figure, the situation
is reversed with respect to fig.~\ref{fig:T2}(a), since now the
dominant contribution comes from the pole B, instead of the pole A
in the former case
and the interference is now constructive.
 We think that the study of this reaction should
in any case bring some additional information to the one done in
ref.~\cite{Alavi-Harati:2001ic} and could shed some light into
the issue of the two $K_1(1270)$ resonances.

\section{Conclusions}

We have evaluated the radiative decays of the low-lying 
axial-vector resonances into a pseudoscalar meson and a photon. For
that purpose, we have extended a previous model originally devoted
to the charged $a_1$ and $b_1$ radiative decay. In our model, the
axial-vector resonances appear as dynamically generated through the
interaction of a vector and a pseudoscalar meson, in the sense that
they appear as poles in unphysical Riemann sheets of the scattering
amplitudes without the need to include them as explicit degrees of
freedom. Within this model the couplings of the axial-vector mesons
to the different $VP$ channels can be easily obtained, even the
relative signs which are crucial in the interferences of the
present work. We evaluate the radiative decay widths by allowing the
photon to be emitted from the decaying $VP$ product both at tree
and one loop level contribution.

We make predictions for all these radiative decay widths and show
that the final results are strongly affected by non-trivial
interferences between different mechanisms, which are under control
thanks to the knowledge of the couplings provided by the  underlying
unitary theory that generates dynamically the axial-vector resonances.
This makes the final results to span a wide range of radiative
widths from $0$ to $\sim 1$~MeV.

We have devoted special attention to the  $K^0_1(1270)\to
K^0\gamma$ decay for which there is only one experimental datum. 
In the underlying model of the present work this resonance has a
double pole structure. We have discussed that, should this be the
actual case in nature, it would have deep consequences in the
experimental result since, in the experimental analysis, the usual
one pole structure of the $K_1(1270)$ was considered. We have also
proposed a 
related experiment using the Primakoff method with $\rho K$ in the
final state instead of the $K^*\pi$ of 
ref.~\cite{Alavi-Harati:2001ic} in order to bring extra
information on the issue of the two pole structure of the
$K_1(1270)$ resonance.
 We argue that the
result obtained for the radiative decay in both Primakoff
experiments should be the same 
if there is only one pole. Yet, if there are two poles, the single
pole analysis is inappropriate and would most probably lead to
different results for the $K_1(1270)$ radiative width in the two
experiments.

  Further experimental measurements of the radiative decay widths
of the  axial-vector resonances 
 would be welcome to shed more light on the
nature of these resonances.

\section*{Acknowledgments}  

This work is partly supported by DGICYT contract number
FIS2006-03438, the Generalitat Valenciana and  the JSPS-CSIC
collaboration agreement no. 2005JP0002, and Grant for Scientific
Research of JSPS No.188661.
One of the authors (H.N.) is supported by JSPS Research Fellowship for
Young Scientists. 
This research is  part of
the EU Integrated Infrastructure Initiative  Hadron Physics Project
under  contract number RII3-CT-2004-506078.
H.N. would like to express her thanks to
Satoru Hirenzaki who supported her stay in Valencia.


\end{document}